# A Survey on Dynamic Job Scheduling in Grid Environment Based on Heuristic Algorithms

**D. Thilagavathi**
Assistant Professor
Department of Computer Science
Nallamuthu Gounder Mahalingam College,
Pollachi, Coimbatore,
Tamilnadu, India

**Dr. Antony Selvadoss Thanamani**
Associate Professor and Head
Department of Computer Science
Nallamuthu Gounder Mahalingam College,
Pollachi, Coimbatore,
Tamilnadu, India

**ABSTRACT**

Computational Grids are a new trend in distributed computing systems. They allow the sharing of geographically distributed resources in an efficient way, extending the boundaries of what we perceive as distributed computing. Various sciences can benefit from the use of grids to solve CPU-intensive problems, creating potential benefits to the entire society. Job scheduling is an integrated part of parallel and distributed computing. It allows selecting correct match of resource for a particular job and thus increases the job throughput and utilization of resources. Job should be scheduled in an automatic way to make the system more reliable, accessible and less sensitive to subsystem failures. This paper provides a survey on various heuristic algorithms, used for scheduling in grid.

## 1. Introduction

The term grid computing refers to the combination of computer resources from multiple administrative domains to reach a common goal. The Grid can be thought of as a distributed system with non-interactive workloads that involve a large number of files. What distinguishes grid computing from conventional high performance computing systems such as cluster computing is that grids tend to be more loosely coupled, heterogeneous, and geographically dispersed.

Grid systems are classified into two categories: compute and data Grids. In compute Grids the main resource that is being managed by the resource management system is compute cycles (i.e. processors), while in data Grids the focus is to manage data distributed over geographical locations. The architecture and the services provided by the resource management system are affected by the type of Grid system it is deployed in. Resources which are to be managed could be hardware (computation cycle, network bandwidth and data stores) or software resources (applications).

A Grid scheduler (or broker) must make resource selection decisions in an environment where it has no control over the local resources, the resources are distributed, and information about the systems is often limited or dated. Here, schedulers are responsible for job management like allocating resources needed by particular job, partitioning of the job to run the job in the parallel manner in parallel processing environment, management of data, event correlation and service-level management capabilities. A Grid scheduler is different from local scheduler in that a local scheduler only manages a single site or cluster and usually owns the resource.As heuristic techniques are increasingly being used for solving optimization problems, they have proven themselves as a good candidate in this area. This can be inferred by recent research in the area.

A large number of heuristic algorithms have been designed to schedule tasks to machines on grid computing systems. The commonly used algorithms are Opportunistic Load Balancing (OLB), Max-min, Min-min, Minimum Execution

Time (MET), Minimum Completion Time (MCT), User Directed Assignment (UDA), Genetic simulated annealing (GSA), Tabu search, simulated annealing (SA), genetic algorithm (GA), ant colony optimization (ACO) and particle swarm optimization (PSO). In this survey some of the above mentioned methods will be discussed.

The rest of the paper is organized as follows: Section 2 describes the grid scheduling process Section 3 discusses the various heuristic Algorithms for Job Scheduling in Grid and finally Section 4 concludes this paper.

## Grid Scheduling process

Grid scheduling process [3] involves three main phases: resource discovery, which generates a list of potential resources; information gathering about those resources and selection of a best set; and job execution, which includes file staging and cleanup. These phases, and the steps that make them up, are shown in Figure 1.

| Phase One-Resource Discovery | Phase Three-Job Execution |
|---|---|
| 1. Authorization Filtering | 6. Advance Reservation |
| 2. Application Definition | 7. Job Submission |
| 3. Min Requirement Filtering | 8. Preparation Tasks |
| **Phase Two-System Selection** | 9. Monitoring Progress |
| 4. Information gathering | 10. Job Completion |
| 5. System selection | 11. Clean-up Tasks |

**Fig 1. Three-phase plan for Grid scheduling.**

**Resource Discovery** Resource discovery involves the user selecting a set of resources to investigate in more detail. At the beginning of this phase, the potential set of resources is empty set and at the end of this phase, the potential set of resources is some set that has passed a minimal feasibility requirement. Most users do this in three steps namely:

**Authorization filtering:** It is generally assumed that a user will know which resources he has access to in terms of basic services. At the end of this step the user will have a list of machines or resources to which he has access. *Application Requirement Definition:* In order to proceed in resource discovery, the user must be able to specify some minimal set of job requirements in order to further filter the set of feasible resources. The set of possible job requirements can be very broad and vary significantly between jobs. It may include static details and Dynamic details. *Minimal Requirement Filtering:* The user generally does this step by going through the list of resources and eliminating the ones that do not meet the job requirements as much as they are known. It could also be combined with the gathering more detailed information about each resource.

**System Selection** In this step a single resource (or single resource set) must be selected on which to schedule the job. This is generally done in two steps: Gathering Information (QUERY)*:* In order to make the best possible resource match, a user needs to gather dynamic information about the resources in question. **Select the system(s) to run on:** Given the information gathered by the previous step, a decision of which resource (or set of resources) should the user submit a job is made in this step.

Job Execution The third phase of scheduling is running a job. This involves a number of steps: Make an Advance Reservation (Optional): It may be the case that to make the best use of a given system, part or all of the resources will have to be reserved in advance. Submit Job to Resources*:* Once resources are chosen the application must be submitted to resources. Preparation Tasks*:* The preparation stage may involve setup, claiming a reservation, or other actions needed to prepare the resource to run the application. Monitor Progress: Depending on the application and its running time, users may monitor the progress of their application. Find out if Job is done*:* When the job is finished, the user needs to be notified.

Completion Tasks**:** After a job is run, the user may need to retrieve files from that resource in order to do analysis on the results, break down the environment and remove temporary settings etc.

3. **Heuristic Algorithms for Job Scheduling in Grid**

Scheduling in Grid though has been intensively studied only during the recent year, there exists a great variety of the algorithms for scheduling in Grid. This section discusses some of the research works on the algorithms for scheduling in grid used.

Ant Colony Optimization (ACO) is a heuristic algorithm with efficient local search for combinatorial problems. ACO imitates the behavior of real ant colonies in nature to search for food and to connect to each other by pheromone laid on paths travelled. Many researches use ACO to solve NP-hard problems such as travelling salesman problem, graph colouring problem, vehicle routing problem, and so on. In paper [4] Ruay-Shiung Chang et al suggests modified ant algorithm as Balanced ACO (BACO) algorithm which reduces makespan time and also tried to balance the entire system load. This work was implemented in the Taiwan UniGrid Platform. The BACO algorithm selects a resource for submitting the request (job) by finding the largest entry in the Pheromone Indicator (PI) matrix among the available jobs to be executed. This work was carried for independent jobs and not for workflow jobs.

In paper [5], the author addresses a new scheduling algorithm called RASA. Here it takes advantages of MIN-MIN and MAX-MIN algorithm and tries to avoid their drawbacks. The algorithm builds a matrix C where $C_{ij}$ represents the completion time of the task $T_i$ on the resource $R_j$. If the number of available resources is odd, the Min-min strategy is applied to assign the first task, otherwise the Max-min strategy is applied. The remaining tasks are assigned to their appropriate resources by one of the two strategies, alternatively. Min-min and Max-min algorithms are applicable in small scale distributed systems. Whereas RASA can be applied to large scale distributed systems.

PSO is a population-based search algorithm based on the simulation of the social behavior of bird flocking and fish schooling. HU Xu-Huai et al [6] proposes an Immune Particle Swarm Optimization (IPSO) algorithm. The basic idea of the IPSO is to record the particles with a higher fitness in the evoluting process, and make the new particles which satisfy neither the assumption nor the constraint condition replaced by the recorded ones. In addition, immune regulation should be done to maintain the species diversity while it decreases. This paper mainly discusses the independent task scheduling. Experiments show that the PSO algorithm has the best integrate performance. Whatever we consider the scheduling creating time, the makespan and the mean response time, it all has good performance.

Genetic algorithm may be used to solve optimization problems by mimicking the genetic process of biological organisms. In [7] authors investigate the job scheduling algorithm in grid environments as an optimization problem. This paper gives an improved genetic algorithm with limited number of iteration to schedule the independent tasks onto Grid computing resources. The evolutionary process is modified to speed up convergence as a result of shortening the search time, at the same time obtaining a feasible scheduling solution. The scheduling creating process in GA algorithm costs the longest time.

In paper [8] author proposes a modified ant algorithm for Grid scheduling problem that is combined with local search. The proposed ant algorithm takes into consideration the free time of the resources and the execution time of the jobs to achieve better resource utilization and better scheduling. The ant algorithm output is sent to local search technique which helps to reduce the overall make span further.

In paper [9] the author uses Algorithm for Task scheduling using PSO with improved job grouping is proposed and implemented. Number of user jobs and the number of resources available are submitted to the scheduler. User jobs are also submitted to the scheduler. Total processing capability of all the available resources is calculated (in MIPS). Total length (in MI) of all the tasks is calculated (Sum of the length of all the submitted tasks). The percentage of the processing capability of a resource on the total processing capability of all the resources is calculated. Using this percentage, the processing capability of a resource based on the total length of all tasks to be scheduled is calculated. By this way the jobs are allocated to the available resources not uniformly, but the utilization of resources will be increased. The scheduler groups the jobs according to the calculated processing capability. The new job group is scheduled to execute in the available resources.

This process of grouping and scheduling is repeated until all the user jobs are grouped and assigned to selected grid resources. The grouped jobs are sent to the selected resources for execution. After execution the job groups will be sent back to the scheduler. It collects all the processed job groups. Due to job grouping this approach optimizes computation/communication ratio and enhances utilization of resources.In paper [10], author presents a framework which combines the Fuzzy C-Mean clustering with an Ant Colony Optimization (ACO) algorithm to improve the scheduling decision when the grid is heterogeneous. In this model, the Fuzzy C-Mean algorithm classifies the jobs into appropriate classes, and the ACO algorithm maps the jobs to the appropriate resources.

P. Mathiyalagan et al [11], proposes ant colony algorithm improved by enhancing pheromone updating rule such that it schedules the tasks efficiently and better resource utilization. The improved pheromone updating rule is given by

$$\tau_{ij}(t)new = [\{\rho + (1-\rho/1+\rho)\} * \tau_{ij}(t)old] + [\{\rho - (\rho/1+\rho)\} * \Delta\tau_{ij}(t)] \quad (1)$$

Where, $\tau_{ij}(t)$ →Trail intensity of the edge (i,j).

$\rho$ →Evaporation rate(0.4).

$\Delta\tau_{ij}(t)$ →Additional pheromone when job moves from scheduler to resource(0.2).

Hesam Izakian et al [12], represents a discrete Particle Swarm Optimization (DPSO) approach for grid job scheduling. Particles need to be designed to present a sequence of jobs in available grid machines. Also the velocity has to be redefined. Equation (2) is used for updating the velocity matrix and then Equation (3) for position matrix of each particle.

$$V_k^{(t+1)}(i, j) = V_k^t(i,j) + c_1r_1(pbest^t_k(i, j) - X^t_k(i, j)) + c_2r_2(nbest^t_k(i, j) - X^t_k(i, j)) \quad (2)$$

$$X_k^{(t+1)}(i, j) = \begin{cases} 1 & \text{if } (V_k^{(t+1)}(i, j) = \max\{V_k^{(t+1)}(i, j)\}), \forall i \in \{1, 2, ..., m\} \\ 0 & \text{otherwise} \end{cases} \quad (3)$$

In Equation (2) $V_k^t(i, j)$ is the element in $i^{th}$ row and $j^{th}$ column of the $k^{th}$ velocity matrix in $t^{th}$ time step of the algorithm and $X^t_k(i,j)$ denotes the element in $i^{th}$ row and $j^{th}$ column of the $k^{th}$ position matrix in $t^{th}$ time step. Equation (3) means that in each column of position matrix, value 1 is assigned to the element whose corresponding element in velocity matrix has the maximum value in its corresponding column. If in a column of velocity matrix there is more than one element with maximum value, then one of these elements is selected randomly and 1 assigned to its corresponding element in the position matrix. DPSO is proved to obtain highest fitness values. Also ACO and Fuzzy Particle Swarm Optimization (FPSO) are in the next ranks.

Mohd Kamir Yusof et al [13], focuses on Tabu Search (TS) algorithm, a scheduling technique in grid computing was tested and evaluated on universal datasets using GridSim tool. The results indicate performance of tardiness is directly related to number of machines up to certain number of resources. The basic principle of TS is to pursue Local Search (LS) whenever it encounters a local optimum by allowing non-improving moves; cycling back to previously visited solutions is prevented by the use of memories, called Tabu lists, that record the recent history of the search, a key idea that can be linked to Artificial Intelligent concepts. TS deal with various techniques for making the search more effective. These include method for exploiting better information that becomes available during search and creating better starting points, as well as more powerful neighborhood operators and parallel search strategies. Another important trend in TS is hybridization, i.e. using TS in conjunction with other solution approaches such as Genetic Algorithm, Lagrangean relaxation, Constraint Programming, column generation and integer programming technique.

In paper [14] Jing Hu et al the author presents an ant colony optimization for grid task scheduling with multiple QoS dimensions (QACO). The proposed algorithm considers five kinds of QoS dimensions: time, reliability, version, security and priority which are transformed to utility as the heuristic information of the algorithm. Here the objective of scheduling is evaluated by total utility. The utility is transformed by QoS requirements through utility functions.

In paper [15] author analyzed Max-Min algorithm and Min-Min algorithm, a modified scheduling algorithm Filter-Min-Min Algorithm was proposed. Max-Min will execute long task first and allows short tasks to be executed concurrently with the long task, resulting better makespan and even better resource utilization rate and load balancing level, compared to Min-Min that executes all short tasks first and then executes the long task. Filter-Min-Min algorithm has tried to remove the disadvantages of both this algorithm.

## 3. Conclusion

In this paper we presented a survey on current trends and applications of Heuristic Algorithms in the area of Grid Scheduling problem. An application which needs high cost resources to be managed and shared required special attention in Grid Environment. In large scale grid, job scheduling provides suitable solutions for managing resources. The grid scheduling problem involves simultaneous optimization of several objectives including completion time, resource utilization, QoS metrics, costs, reliability factors etc. Some heuristic methods discussed in this paper have been used to optimize it and have got some good results.

**Authors Biography**

**Mrs. D. Thilagavathi** received her MCA degree from Bharathidasan University in 2001 and completed her M.Phil. Degree in Computer Science from Bharathiar University in 2005. She is currently pursuing her Ph.D. at the Research Department of Computer Science, NGM College, Pollachi, Bharathiar University, India. Her research interests include Object Oriented Analysis and Design and Grid Computing. She has 11 years of teaching experience. She is now an Assistant Professor and Head in the Department of Computer Technology, NGM College, Pollachi.

**Dr. Antony Selvadoss Thanamani** is presently working as an Associate Professor and Head, Research Department of Computer Science, NGM College, Pollachi, Coimbatore, India. He has published many papers in international/national journals and written many books. His areas of interest include E-Learning, Software Engineering, Data Mining, Networking, Parallel and Distributed Computing. He has a credit of 25 years of teaching and research experience.